\documentclass{article}[14pt,article]
\usepackage{amssymb,amsmath,amsthm}
\def\la{\langle}
\def\ra{\rangle}
\def\ot{\otimes}

\begin{document}

\title{ A note on extremal  states of composite quantum systems with fixed marginals} 

\author{S. Kanmani, \\ Materials Physics Division, IGCAR, \\ Kalpakkam-603 102, India.\\ kanmani@igcar.gov.in \\}

\maketitle

\abstract{ 
An extremal element of the convex set of composite quantum states in $M_2\otimes M_3 $ whose marginals are all normalised identities
has been constructed. It is found to be a  mixed state and  is  entangled as well. }\\



\section{Introduction}  We work with Hilbert spaces that are finite dimensional and are over complex numbers. 
 The  state of a finite level quantum mechanical system associated with an n-dimensional Hilbert
space $H$ is a positive semidefinite matrix $\rho$  of unit trace and is an element of $M_n(C)$. Such an operator is called a
density operator or density matrix.   Equivalently, $\rho $ is a self-adjoint matrix
with non-negative eigenvalues, such that the  sum of the eigenvalues is 1. Physical observables like energy or momentum are
represented by self-adjoint operators.  Then the expectation value of an observable $A \in M_n(C) $, when the physical
system is in the state $\rho$ is given as Tr$(A\rho)$. The set of all states of $H$ is a compact convex set whose extreme points are the rank-one self-adjoint projection operators, called the pure states. A pure state can also be characterised by a unit vector in the range of the corresponding density operator. The states that are not extremal are called mixed states.\\

Let $H_1$ be a complex Hilbert space that represents a finite  level quantum system $S_1$. Similarly, let $H_2
$ be a complex Hilbert space of another finite level quantum system  $S_2$. Then the coupled or composite
quantum  system of  $S_1$ and $S_2$ put together and treated as a single quantum system $S_{12}$, is
represented by the Hilbert space, $H_1\ot H_2$, the tensor product of $H_1$ with $H_2$. Any state $\rho \in M_n\ot M_m$ of
the composite system $H_1\ot H_2$ gives rise to marginal states or equivalently reduced states, $ \rho_{1}=
Tr_{H_{2}}\rho \in M_n $ of $H_1$ and  $ \rho_2 = Tr_{H_{1}}\rho \in M_m $ of $H_2$ respectively. Here $ Tr_{H_{i}}$ denotes
the partial trace or relative trace over $H_i$. That is,  $\rho_2= (tr\ot 1)\rho $, and $\rho_1= (1\ot tr)\rho $, where $tr$ denotes the trace operator. A brief  introduction to partial trace  could be found in Bhatia [1]. Any observable $A_1$ on $H_1$ can be  extended to  an observable $A_1\ot \bf 1 $ on $ H_1\ot H_2$. Then  the expectation value Tr$(A_1 \rho_1)$ = Tr $[( A_1\ot \bf
1 ) \rho ]$ for every observable $A_1 \in H_1$, where $\rho_1$ is the reduced state of the composite state $\rho$.
  A similar relation holds for the observables and marginal
states on $H_2$ as well.\\

The state space $ S(H_1\ot H_2) $ of a composite quantum system turns out to be an extremely complicated
convex set. Especially, the interplay between the convex sets  $ S(H_1\ot H_2) $, $ S(H_1) $ and $ S( H_2) $  raises
many interesting questions.\\

\noindent
{\bf Definition-1}  A composite state $\rho $ of $S(H_1 \ot H_2) $ which is an element of  $M_n \ot M_m$ is said to be a separable state if $  \rho = \sum_{i=1}^{n}  \alpha_i \, \rho_{i}^{1} \ot \rho_{_i}^{2} $ where $ \rho_{i}^{1} \in M_n$ and $ \rho_{_i}^{2} \in M_m $ are states in $S(H_1)$ and $S(H_2)$ respectively and $\alpha_i > 0 $ such that  $ \sum_{i=1}^{n} \alpha_i = 1 $. A composite state that is not a separable state is called an entangled state.\\

 Entangled states are  of central interest in quantum information theory. It is known [2] that a state $\rho $  in $M_2\ot M_2 $ or $M_2 \ot M_3$  is separable if and only if
 $(1 \ot trp)\rho $ is a positive definite matrix  in $M_2 \ot M_2 $ or $M_2 \ot M_3$ respectively. Here, $trp$ denotes the transpose operator.
This condition or criterion is known as positive partial transpose (PPT). However, given  a state in $M_n \ot M_m $, where $n,m > 2$  a simple criterion to establish that it is an entangled state is not known and is an important  open problem of quantum information theory.\\

In this context, K.R.Parthasarathy [3] initated a study of convex subset of $S(H_1\ot H_2) $. For two fixed  states
$\rho_1$ and $\rho_2 $ in $S(H_1)$ and $S(H_2)$ respectively, he considered the set of all composite states $\rho $ in $S(H_1 \ot H_2) $ whose reduced states are $\rho_1 $ and $\rho_2$. This set, denoted as  $C[ \rho_1;\rho_2 ]$ is a compact convex set. Focussing on the nature of the extreme points of this convex set, he obtained the following results of theorem-1 and theorem-2 in [3].\\

\noindent
{\bf Theorem-1 } Let $H_1$, $H_2$ be complex finite dimensional Hilbert spaces of dimension $d_1,d_2 $
respectively and $H_1 \ot H_2$ be the coupled or composite system.  Consider, $ C[\rho_1; \rho_2]$   the convex set of all composite states $ \rho $ of $  H_1\ot H_2 $ whose
marginal states in $H_1$ and $H_2$ are $\rho_1 $ and $\rho_2$ respectively. If an element $\rho \in C(\rho_1,
\rho_2)$ is  an extreme point of the convex set $ C[\rho_1;\rho_2]$,  then its rank does not exceed $\sqrt{d_{1}^2 +d_{2}^2 -1}$. In particular, such an extreme element cannot be an invertible matrix.\\

In the physically interesting  special case of $H_1=H_2=C^2 $ and $\rho_1=\rho_2 = \frac{1}{2}{\bf 1_2}$, he obtained the following result of theorem-2, which completely characterises the extreme points of $C(\frac{1}{2} {\bf 1_2} ,\frac{1}{2} {\bf 1_2} )$.
\vspace{.1in}

\noindent
{\bf Definition-2 } A  composite pure state  $\rho_{\tau} $ in $M_2 \ot M_2$  is called a maximally entangled state if it is a rank-one self-adjoint projection operator,
whose range is spanned by a non-elementary tensor  $\tau = e_1 \ot f_1 + e_2\ot f_2 \in C^2 \otimes C^2 $ where $e_1 =(1,0), e_2=(0,1) $ is the standard basis of $ C^2$ and $ \{f_1,f_2 \} $ is any arbitrary orthonormal basis of $ C^2$.\\

\noindent
{\bf Theorem-2 } Let $ H_1\ot H_2 =C^2 \ot C^2 $ and $C[\frac{1}{2} {\bf 1_2} ;\frac{1}{2} {\bf 1_2} ]$ be the
convex set of composite states with marginals $ \rho_1= \frac{1}{2} {\bf 1_2} $ in $ H_1=C^2$ and $ \rho_2=
\frac{1}{2} {\bf 1_2} $  in $ H_2=C^2$. Then, the composite state $\rho $ is an
extreme point of $C[\frac{1}{2} {\bf 1_2} ;\frac{1}{2} {\bf 1_2} ]$ if and only if $\rho$ is a maximally entangled pure state as defined above.\\

It was suggested by G.L.Price and S. Sakai [4], that the above case is probably true for every $C[\frac{1}{n} {\bf 1_n} ;\frac{1}{m} {\bf 1_m} ]$. However,
it has been shown by Oliver Rudolph [5], and  H.Ohno [6] by  constructing 
concrete examples that there are extreme elements of  $C[\frac{1}{3} {\bf 1_3} ;\frac{1}{3} {\bf 1_3} ]$, and 
$C[\frac{1}{4} {\bf 1_4} ;\frac{1}{4} {\bf 1_4} ]$, that are not  pure states. The work of 
L.J. Landau and R.F.Streater [7] is also relevant and  contains many such  examples.
 A complete characterisation of the  extreme elements of the convex set $C[\frac{1}{n} {\bf 1_n} ;\frac{1}{m} {\bf 1_m} ]$ is not yet known. Apart from theorem-1 not much seems to be known
 of $C[\frac{1}{n} {\bf 1_n} ;\frac{1}{m} {\bf 1_m} ]$, where $n$ is not equal $m$. The smallest in that class is
the coupled system of $M_2 \ot M_3 $, which is also suitable  for the study  of  entanglement  aspects. This is due to fact that the 
PPT (positive partial transposition) criterion,   is a necessary and sufficent condition for separable states only in the case of  $M_2 \ot M_2 $ and  $ M_2 \ot M_3 $.\\ 

In this note, we look at the
case of $ C[\frac{1}{2} {\bf 1_2} ;\frac{1}{3} {\bf 1_3} ]$, and construct a specific extremal state which turns out to be a mixed state that is  entangled.  The tools that we have used are the ones that were developed by  Choi [8], Landau and Streater [7] and  Oliver Rudolph [5]. An introduction to completely positive liner maps is available in the lecture notes of Vern Paulsen [9].\\

\section{Completely  positive linear maps and bipartite composite states}
We construct an extreme element of the convex set $ C[\frac{1}{2} {\bf 1_2} ;\frac{1}{3} {\bf 1_3} ]$. This is made easier, by the
duallity that exists between  of a special class of positive linear map betweem matrix algebras and the set positive semidefinite matrice on the
tensor product of matrix algebras. This  duallity preserves the convexity character of the elements as well. For example, an extremal map corresponds to en extremal positive definite element. To be precise, and specific to this context,  this is 
the well known  Arveson-Choi-Jamiolkovski-de pillis [8-12] correspondence that exist between the set $CP(M_2,M_3) $ of completely positive linear maps from $M_2$ to $M_3$
and the set $PD(M_2\ot M_3)$ of positive semidefinite matrices in $M_2 \ot M_3 $. 
 For a given $\phi \in CP(M_2,M_3) $, we associate the positive semidefinite matrix $\sigma_{\phi}= \sum_{1 \leq i,j \leq 2} E_{ij}\ot \phi(E_{ij}) \in M_2 \ot M_3 $, where $E_{ij}$ represents a matrix
 with 1 in the $(i,j)$ position and zero in every other place.
It is known [8], that a linear map  $ \phi :M_2(C) \rightarrow M_3(C) $ is completly positive if and only if it is of the form $ \phi(A) =\sum_{l=1}^r V^{*}_{l} A
V_{l}$ where $A \in M_2(C)$. Here,  $V_l$ are $ 2 \times 3$ complex matrices and $ V^{*}_{l} $ are their corresponding
conjugate transposes  for $l=1,2,..,r$. It can also be seen, that if $\phi $ is as above its dual $\phi^{*} : M_3(C) \rightarrow M_2(C)$ is of the form $ \phi(A) =\sum_{l=1}^r V_{l} AV^{*}_{l} $, where  $ A \in M_3(C)$. Clearly, $\phi^{*}$ is also a completely positive
linear map.\\

 Let $CP(M_2,M_3; K)$
= $ \{ \phi \in CP(M_2,M_3) : \phi(1_2)= K, $ where $K$ is positive semi-definite matrix  of unit trace in $M_3$ $\}$.  
The duallity or the correspondence $ \phi \in CP(M_2,M_3; K)  \longrightarrow 
\sigma_{\phi} =\sum_{1 \leq i,j \leq 2} E_{ij}\ot \phi(E_{ij})  $ now ensures that $tr(\sigma_{\phi})=1 $ and hence
 $\sigma_{\phi} $ is a composite state
in $M_2 \ot M_3 $. This is because,
$tr(\sigma_{\phi})= tr (\phi(E_{11}))+ tr(\phi(E_{22}))= tr(\phi(1_2))=tr(K )=1 $ by our assumption on the trace of $K$.\\

 Under these conditions, one of the marginal state of $\sigma_{\phi}$  turns out to be $K$ itself.
This is because, $(tr \otimes 1_3)(\sigma_{\phi}) =\sum_{1\leq i,j \leq 2} tr(E_{ij})\,\phi(E_{ij}) =\phi(E_{11}) + \phi(E_{22})=  \phi(1_2)=K $. The last equality follows from our assumption that $ \phi \in  CP(M_2,M_3; K)$, where we demand that $\phi(1_2)= K $.\\

Since, $K \in M_n$ is of unit trace and is positive semi-definite by our assumption, it turns out that   $\sigma_{\phi}$
is a composite state  whose  marginal state, $\rho_2 = K\in M_3(C) $. Thus, the set $CP(M_2,M_3; K)$ specifes, through the above mentioned Arveson-Choi-Jamiolkowski-de pillis
correspondence, a set of composite states in $M_2 \ot M_3 $ whose mariginal state in $M_3$ is $K$.   The correspondence also preserves convexity aspects like  extremality of the elements.\\

The following result is due to Choi [8].\\

\vspace{.1in}

\noindent
{\bf Theorem-3} Let $ \phi : M_n \rightarrow M_m$ be a completely positive linear map such that $\phi (1_2)= K \geq 0 $. Then $\phi$ is extreme in the convex set of $CP(M_n,M_m; K )$ if and only if
$\phi$ admits an expression of the form $ \phi(A)= \sum_{i=1}^{r} V^{*}_i A V_i $ for all $A$ in $ M_n$, where $V_i$ are $n \times m $ matrices, 
$\sum_{i=1}^{r} V_i^{*} V_i=K$ and $ \{ V_i^{*} V_j : 1 \leq i,j \leq r \}  $ is a linearly independent set. \\

The other marginal state \, $\rho_{1} \in M_2(C) $  \, of\,  $\sigma_{\phi} $, when $\phi \in CP(M_n,M_m; K )$ is seen to be
$ \sum_{l=1 }^{r} V_{l}V^{*}_{l} \in M_2(C) $. This is because, $ (1_2 \otimes tr)(\sigma_{\phi})= \sum_{1 \leq i,j \leq 2} E_{ij}
\,tr(\phi(E_{ij})) $= $ \sum_{l=1}^{r} V_{l} V^{*}_{l}$  as $ tr(\phi(E_{ij})) = ( \sum_{l=1}^{r} V_{l} V^{*}_{l})_{ij} $. Note, that 
this  condition can be expressed as $  \phi^{*}(1_3) = \sum_{l=1}^{r} V_{l} V^{*}_{l} $.\\

Let $ CP(M_2,M_3; K, L) = \{ \phi \in CP(M_2,M_3; K,) : \phi^{*}(1_3) = L, $ where $L$ is a positive semi-definite matrix of unit trace in $M_2(C) $ $ \}$. Thus, if $ \phi \in  CP(M_2,M_3; K, L)$ then the marginal states of $ \sigma_{\phi } $ are $L$ and $K$ respectively in $M_2(C)$ and $M_3(C) $. Following result of theorem-4 is is due to Landau and Streater[7]  and Oliver Rudolph [5].\\

 \noindent
{\bf Theorem-4} Let $ \phi : M_n \rightarrow M_m$ and $\phi $ be a  completely positive linear map such that $ \phi(1_2) = K \geq 0 $ and $ \phi^{*} (1_3) = L \geq 0 $. Then $\phi$ is extreme in the convex set of $CP(M_n,M_m; K, L )$ if and only if
$\phi$ admits an expression of the form $ \phi(A)= \sum_{i=1}^{r} V^{*}_i A V_i $ for all $A$ in $ M_n$, where $V_i$ are $n \times m $ matrices, 
such that $\sum_{i=1}^{r} V_i^{*} V_i=K$  \,  \,   $\sum_{i=1}^{r} V_i V^{*}_i=L $ and $ \{  V_i^{*} V_j \oplus V_j V^{*}_{i} : 1 \leq i,j \leq r \}  $ is a independent set of matrices. \\

{\bf Remark. } The above independence means $ \sum_{ 1 \leq i,j \leq r}\, c_{ij} \, V_i^{*} V_j =0  $ and $ \sum_{ 1 \leq i,j \leq r} \,c_{ij}
\, V_j V^{*}_{i} $ =0 implies $ c_{i,j} =0 $ for all $i,j=1,...r .$ \\

\section{ Present work : states in $M_2(C) \otimes M_3(C) $}
In this note we construct an extreme state of the convex set of all composite states in $M_2 \otimes M_3$, whose marginal in $M_2$ is $\frac{1}{2}1_2 $ and  marginal in $M_3$ is $ \frac{1}{3}1_3 $. Using theorem-4 this is equivalent to constructing the $\sigma_{\phi}\in M_2(C) \otimes M_3(C) $, corresponding to the  extreme map $\phi  \in $$CP(M_2,M_3; \frac{1}{3}1_3 $, $ \frac{1}{2}1_2 $ ). This reduces the problem to the construction of a specific  set of $ 2\times 3 $ matrices $\{ V_{i} : i=1,...r  \} $ such that 
$\sum_{i=1}^{r} V_i^{*} V_i=\frac{1}{3}1_3  $ , \, \, $\sum_{i=1}^{r} V_i V^{*}_i= \frac{1}{2}1_2  $ and
 $ \{  V_i^{*} V_j \oplus V_j V^{*}_{i} : 1 \leq i,j \leq r \}  $ is a linearly independent set of matrices. \\

It can verified that the following two matrices $V_1$ and $V_2$  satisfy all the three conditions mentioned above.\\
\noindent
$V_{1}=\left( \begin{array}{ccc}

0 & \frac{1}{\sqrt{6}} & 0 \\
\frac{1}{\sqrt{3} } & 0 & 0  \end{array} \right)$  \\
and\\
$V_{2}=\left( \begin{array}{ccc}

0 & 0 & \frac{1}{\sqrt{3}} \\
0 &\frac{1}{\sqrt{6} } & 0 \end{array} \right) $ .\\

The extremal composite  $\rho $ of $ C[\frac{1}{2}1_2 ;\frac{1}{3}1_3 ] $ that corresponds to the extremal map
 $ \phi (A) = \sum_{l=1}^2 V^{*}_l A V_{l} $, where $V_1$ and $V_2$ are listed above can be computed  and is given below.\\

$\rho =\frac{1}{3}\left( \begin{array}{cccccc}
0& 0 & 0& 0& 0& 0 \\
0 &\frac{1}{2} &0 &\frac{1}{\sqrt{2}}&0 &0 \\
0&0&1 &0 & \frac{1}{\sqrt{2}} &0 \\
0 &\frac{1}{\sqrt{2}}&0 & 1 &0 &0 \\
0&0 &\frac{1}{\sqrt{2}}& 0 &\frac{1}{2} & 0 \\
0& 0 & 0& 0& 0& 0     \end{array} \right) $\\

This extremal state $\rho $ is not a pure state of $M_2(C) \otimes M_3(C) $ as the rank of $\rho$ is two.
This is compatible with the  theorem-1 of Parthasarathy, according to which, the rank of this state should be less than or equal to
 $\sqrt{2^2+3^2 -1}= \sqrt{12} $. \\

 However,
the partial transpose of $\rho $, that is, $ (1_2 \otimes trp) \rho $ is not a  positive semi-definite matrix. This implies, that the state 
$\rho$ is an entangled state. Recall, that the partial transpose of a separable state, a state that is not entangled,
 has to be a positive semi-definite matrix. Thus these results of $M_2(C) \otimes M_3(C) $ are in contrast with the results of K.R.Parthasarathy, in the context of
$M_2 \otimes M_2$,  where an element of $  C[\frac{1}{2}1_2 ;\frac{1}{2}1_2 ] $  is an extremal element if and only if it
is a maximally entangled pure state. It  known [5-7]  that that this result of K.R. Parthasarathy does not extend to
$M_3 \otimes M_3 $ and $M_4 \otimes M_4 $.\\

\vspace{.1in}
\noindent
{\bf References}\\

\begin{item}
\noindent
{\item 1 Rajendra Bhatia, Partial traces and entropy inequalities, Linear algebra applic., 370, 125-132, (2003) }
\noindent
{\item 2 Michal Horodecki, Pawel Horodecki and Ryszard Horodecki, Separability of mixed states: Necessary and sufficient conditions, Phys. Lett. {\bf A 223 }, 1-8, (1996) }
\noindent
{\item 3 K.R.Parthasarathy, Extremal quantum states in coupled quantum systems with fixed marginals, Ann. Inst. H. Poincare, 41, 257-268 (2005). }
{\item 4. G.L.Price and S.Sakai, Extremal marginal tracial states in coupled systems, Operators and matrices, 1,153-163, (2007).}
\noindent
{\item 5. O.Rudolph, On extremal quantum states of composite systems with fixed marginals, J. math. Phys. 45, 4035-4041, (2004). }

{\item 6. H. Ohno, Maximal rank of extremal marginal tracial states, J. Math. Phy. 51, no-9, 092101-092110, (2010) }

{\item 7.L.J.Landau and R.F.Streater, On Birkhoff's theorem for doubly stochastic completly positive maps of matrix algebras, Linear Algebra applic., 193, 107-127, (1993). }

{\item 8. M.D.Choi, Completely positive maps linear maps on complex matrices, Linear algebra applic. 10, 285-290, (1975). }

{\item 9. Vern Paulsen, Lectures on quantum computing and completely positive maps, available
 at www.imsc.res.in/${\sim }$sundar/QC $ {}_{\_}$ Notes.pdf. }
{\item 10. W. Arveson, Subalgebra of $ C^*$ algebras, Acta Math. 123, 141-224, (1969). }

{\item 11. A. Jamiolkowki, Linear transformations which preserve trace and positive semidefiniteness of operators, Rep. Math. phy, 3, 275-278, (1972) }
{\item 12. John de Pillis, Linear transformations which preserve hermitian and positive semidefinite operators, Pacific jour. math, 23, no-1, 129-137, (1967). }

\end{item}

\end{document}